\documentclass[twocolumn,prc,amsmath,showpacs]{revtex4}
\usepackage{amsmath}
\usepackage{color}
\usepackage{graphicx}
\usepackage{amssymb}

\makeatletter

\begin{document}

\title{Long- and short-range correlations in the \textit{ab-initio} no-core
shell model}

\author{Ionel Stetcu}
\altaffiliation{On leave from the National Institute for Physics and Nuclear Engineering ``Horia Hulubei", Bucharest, Romania.}

\author{Bruce R. Barrett}

\affiliation{Department of Physics, University of Arizona, P.O. Box 210081, Tucson,
Arizona 85721}

\author{Petr Navr\'{a}til}

\affiliation{Lawrence Livermore National Laboratory, Livermore, P.O. Box 808,
California 94551}

\author{James P. Vary}

\affiliation{Department of Physics and Astronomy, Iowa State University, Ames,
Iowa 50011}

\date{\today{}}

\begin{abstract}
In the framework of the \textit{ab-initio} no-core shell model (NCSM),
we describe the longitudinal-longitudinal distribution function, part
of the inclusive $(e,e')$ longitudinal response.
In the two-body cluster approximation, we compute the effective operators
consistent with the unitary transformation used to obtain the effective
Hamiltonian. When short-range correlations are probed, the results
display independence from the model space size and length scale. Long-range
correlations are more difficult to model in the NCSM and they can
be described only by increasing the model space or increasing the
cluster size. In order to illustrate the model space independence
for short-range observables, we present results for a large set of
model spaces for $^{4}$He, and in $0-4\hbar\Omega$ model spaces
for $^{12}$C.
\end{abstract}


\maketitle
Atomic nuclei are the result of a delicate interplay between short-
and long-range correlations among the nucleons, which makes their
theoretical description rather challenging. For light nuclei, very
successful methods, such as Green's function Monte Carlo \cite{GFMC},
hyperspherical harmonics \cite{hh}, and the no-core shell model (NCSM)
\cite{spectr6,c12lett,c12,navratil2003,H3}, have been developed recently.
They allow an \textit{ab-initio} description of nuclear properties,
the only ingredients being realistic nucleon-nucleon (NN) interactions,
which describe the experimental phaseshifts with high accuracy, and
theoretical three-nucleon forces. 

In the \textit{ab initio} NCSM, one starts with a realistic NN interaction
(theoretical three-body forces can also be used, but we will not discuss
this case here) and performs a unitary transformation \cite{okubo,LS80,UMOA}
to a model space, which allows an exact diagonalization in a finite
many-body space, defined by the number of excitations above a mean-field-like
configuration. Details of the procedure are available to the interested
reader in previous publications \cite{spectr6,c12lett,c12,navratil2003}.
Recently, we have extended the same procedure from the Hamiltonian
to general one- and two-body operators \cite{effop,brucefest,navratil1997}.
The effect of the procedure is to reduce the dependence of the observables
upon the model space and harmonic oscillator (HO) frequency, and,
in the lowest approximation, it has proven to be effective only for
short-range operators \cite{effop}.

Using the unitary transformation approach \cite{okubo,UMOA,LS80},
we obtain the following expression for the effective operators \cite{navratil1993,okubo}\begin{equation}
P{\cal O}P=\frac{P+P\omega^{\dagger}Q}{\sqrt{P+\omega^{\dagger}\omega}}O\frac{P+Q\omega P}{\sqrt{P+\omega^{\dagger}\omega}},\label{eq:effOp}\end{equation}
 where the transformation operator $\omega$ satisfies the condition
$Q\omega P=\omega$, with $P$ and $Q$ the projector operators in
the model and complementary spaces, respectively, and $O$ is the
bare general operator, which acts in the entire space. 

In principle, the transformation operator $\omega$ can be computed
using a finite set of eigenvectors in the full space, making use of
the overlap of the full space eigenvectors with the basis states in
the $P$ and $Q$ spaces \cite{c12lett,c12}, \textit{i.e.,}\begin{equation}
\langle\alpha_{Q}|\omega|\alpha_{P}\rangle=\sum_{k\in{\mathcal{K}}}\langle\alpha_{Q}|k\rangle\langle\tilde{k}|\alpha_{P}\rangle.\label{omega}\end{equation}
In the last equation, $|\alpha_{P}\rangle$ and $|\alpha_{Q}\rangle$
are the basis states in the P and Q spaces, respectively. The summation
runs over a finite subset, ${\cal K}$, of eigenvectors in the full
space, and the tilde stands for the inverse of the overlap matrix,
\textit{i.e.}, $\sum_{\alpha_{P}}\langle k'|\alpha_{P}\rangle\langle\alpha_{P}|\tilde{k}\rangle=\delta_{kk'}$.

Equation (\ref{omega}) shows that in order to obtain the transformation
operator $\omega$ one needs the solution to the initial $A$-body
problem. This makes its application impractical, unless we use approximations.
In the simplest approximation, the transformation operator $\omega$
and, therefore, the effective interaction are obtained in the relative
system of two particles, in a large HO basis. The $Q$ space is chosen
to be a few hundred $\hbar\Omega$ excitations in order to obtain
an exact solution to the two-body Schr\"{ o}dinger equation. Due
to the rotational symmetry, we formulate the problem in two-nucleon
channels with good total spin $s$, total angular momentum $j$, and
isospin $t$, reducing drastically the dimensions involved, when performing
the summation over the states in the $Q$ space in Eq. (\ref{eq:effOp}).
The same procedure can be applied to operators that can be analytically
expressed in terms of relative and center-of-mass coordinates of pairs.
However, note that in the case of non-scalar operators calculations
performed with Eq. (\ref{eq:effOp}) become more difficult, because
such operators can in general couple different channels. As expected,
since the transformation operator is a scalar, this procedure changes
neither the character nor the rank of the bare operator. By construction,
keeping the cluster approximation fixed (in this case, the two-body
cluster) and increasing the model space decreases the effect of the
renormalization. This was demonstrated in earlier papers in the case
of $^{3}$H for the ground-state energy \cite{H3}, and in this paper
we will also demonstrate this by comparisons between the effective
and bare results for an observable related to the Coulomb sum rule. 

The inclusive $(e,e')$ longitudinal data presents one of the clearest
experimental signatures for short-range correlations in the wave-function
of the ground state, at least for light nuclei. The Coulomb sum rule\begin{equation}
S_{L}(q)=\frac{1}{Z}\int_{\omega_{el}}^{\infty}d\omega S_{L}(q,\omega)\label{eq:CSR}\end{equation}
is the total integrated strength measured in electron scattering.
In Eq. (\ref{eq:CSR}), $S_{L}(q,\omega)=R(q,\omega)/|G_{E,p}(q,\omega)|^{2}$,
with $R(q,\omega)$ the longitudinal response function and $G_{E,p}(q,\omega)$
the proton electric form factor, while $\omega_{el}$ is the energy
of the recoiling $A$-nucleon system with $Z$ protons. The Coulomb
sum rule $S_{L}(q),$ which is related to the Fourier transform of
the proton-proton distribution function \cite{ppdist}, can be expressed
as \cite{ppdist2}\begin{eqnarray*}
S_{L}(q) & = & \frac{1}{Z}\langle g.s.|\rho_{L}^{\dagger}(\mathbf{q})\rho_{L}(\mathbf{q})|g.s.\rangle-\frac{1}{Z}|\langle g.s.|\rho_{L}(\mathbf{q})|g.s.\rangle|^{2}\\
 &  & \equiv1+\rho_{LL}(q)-ZF_{L}(q)/G_{E,p}(q,\omega_{el}),\end{eqnarray*}
where $F_{L}(q)$ is the longitudinal form factor. If one neglects
the relativistic corrections and two-body currents, $\rho_{L}(\mathbf{q})$
is simply the charge operator\[
\rho_{L}(\mathbf{q})=\frac{1}{2}\sum_{i=1}^{A}\exp(i\mathbf{q}\cdot\mathbf{r}_{i})(1+\tau_{z,i}).\]
 Consequently, the longitudinal-longitudinal distribution function
becomes \cite{ppdist2}\[
\rho_{LL}(q)=\frac{1}{4Z}\sum_{i\neq j}\langle g.s.|j_{0}(q|\mathbf{r}_{i}-\mathbf{r}_{j}|)(1+\tau_{z,i})(1+\tau_{z,j})|g.s.\rangle.\]

\begin{figure}[!t]
\includegraphics[%
  clip,
  scale=0.4]{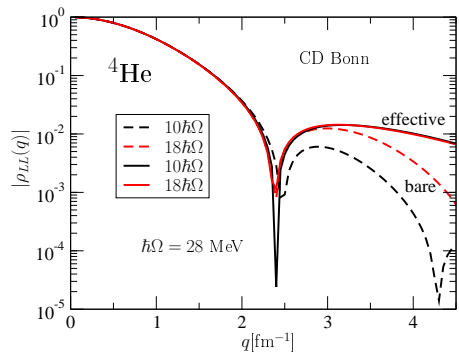}

\caption{(Color online) The longitudinal-longitudinal distribution function
$\rho_{LL}(q)$ in $^{4}$He for two model spaces ($10\hbar\Omega$
and $18\hbar\Omega$) and fixed frequency $\hbar\Omega=28$ MeV, using
bare (dashed curves) and effective (continuous curves) operators.
As discussed in the text, the results obtained with effective operators
are almost indistinguishable.\label{fig:He4}}
\end{figure}

\begin{figure}[!t]
\includegraphics[%
  clip,
  scale=0.45]{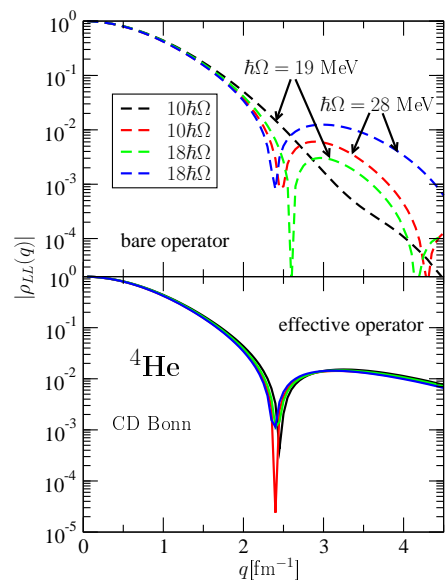}

\caption{(Color online) Longitudinal-longitudinal distribution function, using
bare (upper panel) and effective (lower panel) operators. We used
two different HO frequencies, 19 MeV and 28 MeV, and two model spaces,
$10\hbar\Omega$ and $18\hbar\Omega$.\label{fig:He4m}}
\end{figure}

\begin{figure}[!t]
\includegraphics[%
  clip,
  scale=0.5]{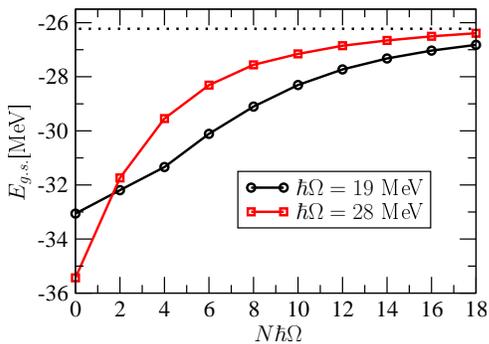}

\caption{(Color online) The convergence of the ground-state energy for the
two frequencies used to compute the $\rho_{LL}(q)$. The dotted line
is the exact ground-state energy for the CD-Bonn interaction ($-26.16$
MeV \cite{gsHe4energy}).\label{fig:He4_gs_conv}}
\end{figure}

We present the results for $\rho_{LL}(q)$ for $^{4}$He in Figs.
\ref{fig:He4} and \ref{fig:He4m}. We have limited this investigation
to two-body interactions only, and, in particular, have used the phenomenological
CD-Bonn NN force \cite{bonn}, because it yields reasonable convergence
properties with increasing the size of the model space. Although experimental
data for the longitudinal-longitudinal distribution function exist,
a direct comparison with experiment is not suitable, because we neglect:
on one hand, (i) three-body forces in the model Hamiltonian, and,
on the other hand, (ii) exchange currents and (iii) relativistic corrections
for the charge operator. Nevertheless, the results demonstrate the
behavior of short- and long-range operators within the framework of
the NCSM.

Using a Gaussian operator of variable range, we have shown previously
how the renormalization of this two-body operator depends upon its
range \cite{effop}. We found that a short-range two-body operator
is renormalized accurately at the two-body cluster level, while a
long-range operator is weakly renormalized. The same behavior can
also be inferred from Fig. \ref{fig:He4}, where we present $\rho_{LL}(q)$
calculated in two model spaces for a fixed frequency, using both bare
and effective operators. Thus, at large momentum transfer, the use
of effective operators produces model-space independent results. Even
in small model spaces we obtain good results, although Fig. \ref{fig:He4_gs_conv}
shows that the ground-state wave function is not fully converged in
such small spaces, since the ground-state enegy is not converged to
the exact value. In particular, for $N_{max}=10$ (or $10\hbar\Omega,$
in terms of allowed excitations beyond the lowest configuration) the
ground-state energy is $-28.30$ MeV for $\hbar\Omega=19$ MeV and
$-27.56$ MeV for $\hbar\Omega=28$ MeV, compared to the exact $^{4}$He
CD-Bonn ground-state energy of $-26.16$ MeV \cite{gsHe4energy}.
The complete convergence of the ground-state energy can be obtained
within the NCSM, as demonstrated, e.g., in Fig. 1 of Ref. \cite{Navratil2005}.
As expected, Fig. \ref{fig:He4} shows that in the large model space
the renormalization is weaker, \textit{i.e.,} there is less need for
renormalization, so that the value obtained with the bare operator
is similar to the value obtained with the renormalized operator. Thus,
for $q\lesssim3$ fm$^{-1}$ one cannot distinguish between the results
obtained using bare operators in the $18\hbar\Omega$ model space
for HO frequency $\hbar\Omega=28$ MeV and the ones obtained using
effective operators. For momenta $q>3$ fm$^{-1}$ the results using
the bare operator deviate from the renormalized values, because the
short-range correlations induced by the interactions are cast into
the effective interaction. If one wants to account for short-range
correlations using a bare short-range operator, one has to increase
the model space, so that the effect of the short-range renormalization
is negligible. However, such a scheme would require a vast number
of $\hbar\Omega$ excitations to obtain a convergent result. 

In Fig. \ref{fig:He4m}, we present $\rho_{LL}(q)$ calculated in
$10\hbar\Omega$ and $18\hbar\Omega$ model spaces, with HO frequencies
of 19 and 28 MeV. In the upper panel we show the results obtained
using bare operators. In this case, the values are spread over orders
of magnitude. In contrast, the lower panel demonstrate independence
of \textit{both} model space and frequency, when using the appropriate
effective operator, although the ground-state energy is not converged,
as illustrated in Fig. \ref{fig:He4_gs_conv}. Moreover, Fig. \ref{fig:He4m}
shows that the convergence depends strongly upon the HO frequency,
when using bare operators. Thus, the results obtained with the bare
operator in $18\hbar\Omega$ with $\hbar\Omega=19$ MeV are far from
the results using effective operators; moreover, this curve shows
a second minimum around $q\simeq4.25$ fm$^{-1},$ whereas the converged
results are almost flat and several orders of magnitude larger for
this value of the momentum transfer. In contrast, even if still significantly
different from the converged values, the results for $\hbar\Omega=28$
MeV are closer to the ones obtained with effective operators.

\begin{figure}[!t]
\includegraphics[%
  clip,
  scale=0.52]{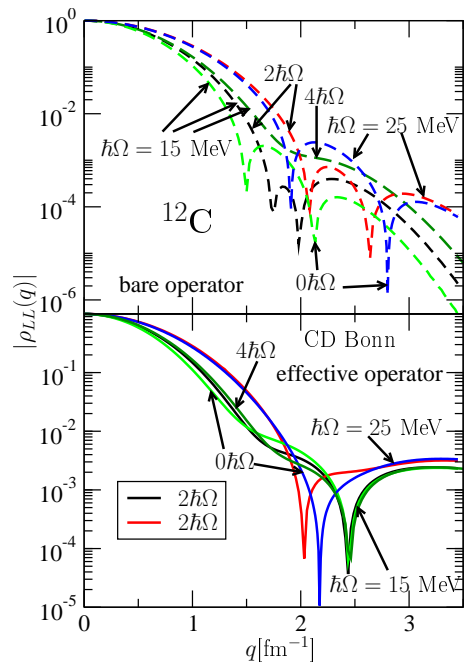}

\caption{(Color online) Longitudinal-longitudinal distribution function for
different model spaces and frequencies in $^{12}$C, using bare (upper
panel) and effective (lower panel) operators. For the $4\hbar\Omega$
model space, we show only the results using $\hbar\Omega=15$ MeV.
While the frequency dependence is not completely removed, the use
of effective operators produces indistinguishable results at large
$q$ for different model spaces, as discussed in the text. \label{fig:C12}}
\end{figure}

One can better observe the influence of the frequency and model space
in Fig. \ref{fig:C12}, where we present the results for $^{12}$C.
For $^{12}$C, unlike the case of $^{4}$He, where we have used a
Jacobi-coordinate HO basis (see, e.g., Ref. \cite{H3}), the investigation
was performed using the Many-Fermion Dynamics code \cite{MFD}, which
employs a Slater determinant basis and becomes much more efficient
for $A>5$ than a Jacobi-coordinate approach. In this case, the number
of many-body configurations increases very rapidly and one has to
limit the truncation to a smaller model space. We present results
for up to $4\hbar\Omega$ model spaces. Again, note in the upper panel
that the results obtained with the bare operators differ widely in
shape and magnitude. When using effective operators, however, all
curves collapse into the same shape and agree with each other for
$q\gtrsim3$ fm$^{-1}$, as shown in the lower panel. Because the
calculation is not fully converged, the minima still change significantly
with frequency and model space, even when one uses effective operators.
One observes that, although the results at high momentum transfer
are very close together, a small dependence upon the HO frequency
persists. 

In summary, we have investigated the longitudinal-longitudinal distribution
function (part of the Coulomb sum rule) in the framework of the NCSM,
utilizing the two-body cluster approximation. Thus, we have extended
our previous application of the effective operator formalism \cite{brucefest,effop,navratil1997}
to the calculation of an observable that probes the short-range correlations.
We find that even very small model spaces can provide an accurate
description of the short-range observables, \textit{if} effective
operators are employed. This investigation shows that reliable results
can be obtained for short-range operators, even for heavier nuclei,
such as $^{12}$C, for which the $0\hbar\Omega$ results are accurate
at higher $q$. As expected, intermediate- and long-range correlations
can be best described by increasing the size of the model space and/or
by a using higher order cluster approximation.

\bigskip

We thank Andreas Nogga for discussions that have prompted this investigation.
I.S. and B.R.B acknowledge partial support by NFS grants PHY0070858
and PHY0244389. The work was performed in part under the auspices
of the U. S. Department of Energy by the University of California,
Lawrence Livermore National Laboratory under contract No. W-7405-Eng-48.
P.N. received support from LDRD contract 04-ERD-058. J.P.V. acknowledges
partial support by USDOE grant No DE-FG-02-87ER-40371. We thank the
Institute for Nuclear Theory at the University of Washington for its
hospitality and the Department of Energy for partial support during
the development of this work.

\end{document}